\documentclass[twocolumn,prb,superscriptaddress,floatfix]{revtex4-2}
\usepackage{amsmath}
\usepackage{amssymb}
\usepackage{graphicx}
\usepackage{color}
\usepackage{hyperref}
\usepackage{bm}

\begin{document}

\title{Turning a superconducting diode into a Feynman ratchet \\to realize a nanoscale heat engine}
\author{Konstantinos Kontogeorgiou}
\affiliation{Institute for Quantum Information, RWTH Aachen University, 52056 Aachen, Germany}

\author{\.Inan\c{c} Adagideli}
\affiliation{Faculty of Engineering and Natural Sciences, Sabanci University, Orhanli-Tuzla, Istanbul, Turkey}
\affiliation{MESA+ Institute for Nanotechnology, University of Twente, The Netherlands}

\author{G\'eraldine Haack}
\affiliation{Department of Applied Physics \& Geneva Quantum Centre, University of Geneva, 1211 Geneva, Switzerland}

\author{Fabian Hassler}
\affiliation{Institute for Quantum Information, RWTH Aachen University, 52056 Aachen, Germany}
\date{October, 2026}

\begin{abstract}
Feynman's ratchet illustrates how work can be extracted from thermal fluctuations in the presence of two heat baths at different temperatures. While some experimental realizations of the ratchet effect have been studied, most lack miniaturization, which is important for some applications. Here, we propose a physical implementation of Feynman's ratchet based on Josephson junctions exhibiting the superconducting diode effect. This setup functions as a heat engine that generates work solely from a temperature gradient via thermal fluctuations. We analyze its output power and efficiency, and, through the Seebeck coefficient, identify the conditions under which the device operates in reverse, functioning instead as a Peltier cooler.
\end{abstract}

\maketitle

\section{Introduction}

Thermodynamics is fundamentally concerned with heat engines that convert thermal energy into useful work. The second law of thermodynamics prohibits the extraction of work from a single thermal reservoir. When two reservoirs at different temperatures 
$T_1<T_2$ are available, work can be extracted, with an efficiency bounded by the Carnot limit $\eta_C=1-T_1/T_2$. In such a configuration, the engine operates by absorbing heat from the hot reservoir and releasing part of it to the cold one.

Before proceeding further, it is important to distinguish heat engines from passive rectifiers or diodes. A thermal diode can rectify heat currents and allow for heat transport in an asymmetric potential; however, it does not convert thermal energy into work. In contrast, a heat engine transforms a temperature difference into mechanical or electrical work.

More specifically, heat engines can operate through fundamentally different physical mechanisms. In conventional thermoelectric devices, such as those based on the Seebeck effect, a temperature gradient generates a thermodynamic force that drives charge carriers deterministically. The resulting electric current arises from energy-dependent transport properties and can be described within the framework of linear irreversible thermodynamics.

By contrast, a conceptually distinct class of engines is represented by Brownian heat engines, where directed motion results from the rectification of thermal fluctuations in an asymmetric system. A paradigmatic example is Feynman's ratchet~\cite{feynman63}. In this thought experiment, a ratchet mechanism is coupled to two thermal reservoirs that are maintained at different temperatures. Work can be extracted only when the reservoirs are at unequal temperatures. The essential feature of this model is that directed motion emerges from nonequilibrium fluctuations in a spatially asymmetric potential~\cite{Buttiker87,Landauer88}.

To better understand the minimal requirements for such engines, it is useful to discuss the role of degrees of freedom. In stochastic thermodynamics, the dynamics of a system can often be described in terms of a small number of effective variables obeying Langevin equations. If a single degree of freedom is coupled to a single thermal reservoir, detailed balance guarantees that the steady-state probability current vanishes, and no net motion can occur \cite{Reimann2002a, Reimann2002b, Seifert2012}. In other words, thermal fluctuations alone cannot produce directed motion in equilibrium.

Directed motion can only arise under nonequilibrium conditions, such as the presence of multiple reservoirs at different temperatures, external driving, or non-thermal noise sources~\cite{sekimoto1998langevin, astumian1994fluctuation, reimann1997current}. In Feynman's original construction, at least two effective degrees of freedom are required, each coupled to a different thermal reservoir, so that detailed balance is broken by the temperature difference itself. It is precisely this coupling structure that allows Feynman's ratchet to operate as a genuine heat engine rather than as a passive rectifier.

Ratchet effects have been extensively studied in a wide range of systems, including biological motors~\cite{astumian1994fluctuation, Bustamante04, Ai09, Perez10}, photonic setups~\cite{faucheux1995optical, Paneru18, bang2018experimental}, discrete stochastic systems~\cite{jarzynski1999feynman}, molecular dynamics~\cite{Broeck04}, and solid-state systems~\cite{Derenyi98, hanggi2009artificial, Costa23, yuan2022supercurrent, sterck2005three, rollano2021realization}. Of particular relevance to the present work are implementations involving Josephson junctions~\cite{Zapata95, goldobin2001josephson, majer02, Scheer25}, where rectification effects and directed transport have been demonstrated under various nonequilibrium conditions. In such mesoscopic and superconducting platforms, quantum coherence and phase dynamics can further enrich transport properties, and quantum ratchet phenomena have also been explored~\cite{Reimann97, Smirnov08}. Importantly, however, most of these solid-state realizations rely on external AC driving or engineered nonequilibrium noise sources and therefore do not constitute autonomous heat engines powered solely by a temperature difference between two thermal reservoirs.

The mechanism considered here is distinct from the rectification of an applied bias current by superconducting and Josephson diode effects, which stems from different critical currents for opposite bias directions~\cite{ando2020observation,nadeem2023superconducting}. In Josephson systems, this nonreciprocity can arise from asymmetric current-phase relations (CPRs)~\cite{bozkurt2023double,souto22,fominov22,schmid24}. Although such systems can rectify an applied current, this property alone does not make them heat engines. In our device, directed transport instead results from the rectification of thermal fluctuations in an asymmetric Josephson potential, with the temperature difference between two reservoirs providing the nonequilibrium conditions required for work extraction.

Genuine Feynman-type ratchets, in turn, operate as autonomous thermal engines driven exclusively by a temperature gradient between two baths. While several experimental realizations of such thermally driven ratchets have been reported, for instance in photonic systems~\cite{Paneru18, bang2018experimental} and granular gases~\cite{Eshuis10}, scalable solid-state implementations operating purely as two-bath heat engines remain scarce~\cite{pekola07,peltonen11,germanese22}.

In this work, we propose a superconducting circuit realization of Feynman's ratchet. Unlike a thermal diode, our device extracts electrical work from a temperature difference. Unlike conventional thermoelectric engines, its operation relies on the rectification of thermal fluctuations in an asymmetric Josephson potential. The proposed architecture benefits from miniaturization, compatibility with standard electronic circuitry, and operation at cryogenic temperatures.

The manuscript is organized as follows. In Sec.~\ref{sec:setup}, we introduce the setup and relevant parameters. The open-circuit Seebeck voltage generated by a finite temperature gradient is calculated in Sec.~\ref{sec:voltage}, and in Sec.~\ref{sec:Seebeck}, we study the Seebeck coefficient and identify the point where it is maximized. In Sec.~\ref{sec:efficiency}, we analyze the work output and efficiency under a finite load. Next, we discuss the cooling performance in Sec.~\ref{sec:Cooling}. Finally, we summarize our conclusions in Sec.~\ref{sec:conclusion}.

\section{Superconducting diode as Feynman ratchet}\label{sec:setup}

\begin{figure}[tb]
    \centering
    \includegraphics{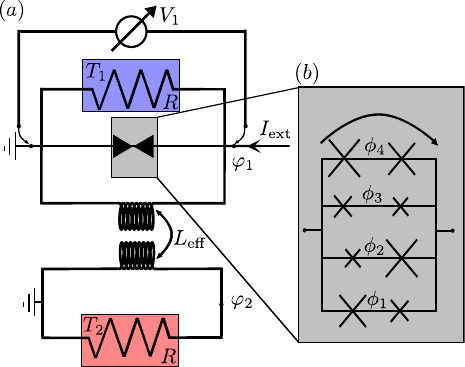}
    \caption{(a) Circuit realizing the Feynman ratchet, with two degrees of freedom $\varphi_1, \varphi_2$. The upper branch, characterized by the phase $\varphi_1$, consists of a superconducting diode (gray box) overdamped by a resistor $R$ at temperature $T_1$ in parallel. The voltage across this branch is denoted by $V_1$. This upper branch is tightly coupled via an inductance $L_\text{eff}$ to the lower branch characterized by the phase $\varphi_2$.  The lower branch contains a resistor $R$ held at a different temperature $T_2$. The voltage of the second branch, in the tight-coupling limit assumed is $V_2 \approx V_1$. The upper branch may be biased by the external current $I_\text{ext}$. (b) For concreteness, we consider an implementation of the superconducting diode as a network of eight Josephson tunnel junctions (four wires) with their corresponding phases $\phi_i$, following Ref.~\cite{bozkurt2023double}. Panel (b) corresponds to the gray box from panel (a).}
    \label{fig:circuit}
\end{figure}

The goal is to realize Feynman's ratchet~\cite{feynman63}, \textit{i.e.}, a passive nonreciprocal element, immersed in a cold bath and coupled to a hot bath, that converts microscopic thermal fluctuations into directed macroscopic motion. To this end, we propose exploiting a superconducting circuit illustrated in Fig.~\ref{fig:circuit}(a). It is made of two branches, the upper one connected to a cold reservoir and the lower one connected to a hot reservoir, both reservoirs being implemented by two resistors $R$ at different temperatures. Both branches interact through a mutual inductance between two coils, see Fig.~\ref{fig:circuit}(a).

As motivated in the introduction, the implementation of a Feynman ratchet requires a nonreciprocal passive element, here a superconducting diode embedded into the upper branch. This diode is characterized by an asymmetric CPR, which produces an asymmetric Josephson potential. In the presence of fluctuations, this asymmetry leads to a directional bias in the phase dynamics. The superconducting phases of the circuit play the role of the macroscopic variables of interest, and their dynamics is governed by the Langevin equations derived below.

\subsection{Dynamics}
By applying Kirchhoff's laws for the two branches of this circuit, we relate the dynamics of the superconducting phases $\varphi_{1,2}$ to the current through the diode $I(\varphi_1)$ and an additional external current $I_\text{ext}$, which may be supplied by an external source and injected into the upper branch. $I_\text{ext}$ drives the system out of the open-circuit regime as we will discuss in Sec.~\ref{sec:efficiency}.  Fluctuating currents generated by the two resistors at temperatures $T_1$ and $T_2$ are taken into account as follows, see details in App.~\ref{app:SDEderivation},
\begin{equation}
\label{eq:system}
\begin{aligned}
    &\frac{\hbar}{2 e R} \dot{\varphi}_1 + I(\varphi_1) + I_\text{ext} + \frac{\hbar}{2 e L_\text{eff}} (\varphi_1 - \varphi_2)  = \delta I_1(t),\\
   & \frac{\hbar}{2 e R} \dot{\varphi}_2  + \frac{\hbar}{2 e L_\text{eff}} (\varphi_2 - \varphi_1) =  \delta I_2(t).
\end{aligned}
\end{equation}
Here, $L_\text{eff}$ is the effective inductance.
The resistors give rise to fluctuating currents, $\delta I_1$ and $\delta I_2$ respectively, characterized by a Johnson-Nyquist noise with Gaussian distribution, $\langle \delta I_j \rangle = 0$ and $\langle \delta I_j(t) \,\delta I_{j'}(t')\rangle = 2 k_B T_j/R \,\delta(t-t') \delta_{jj'}$.
The second Josephson relation relates the voltage in the two branches to the superconducting phases,
\begin{equation}
\label{eq:volphase}
    V_j = \frac{\hbar}{2 e} \dot{\varphi}_j\,, \quad j=1,2\,.
\end{equation}
In the following, we assume a small effective inductance between the two branches, allowing us to make the approximation $V_1 \approx V_2$, akin to a transformer with unit turns ratio. Here, let us comment on the role of the temperatures and the asymmetric CPR. In the model we consider in this work, a finite voltage will arise under two necessary conditions: first, the asymmetry of the CPR, which ensures a preferred direction for the current, and second, a non-zero temperature gradient set by $T_1 \neq T_2$, which ensures a finite current flow.

From now on, for convenience, we define dimensionless variables that will further simplify our expressions. The dimensionless coupling induced by the mutual inductance is given by
\begin{equation}
    \varkappa =\frac{t_0 R}{L_{\text{eff}}},
\end{equation}
with $t_0 = (\hbar/(2 e))^2 /(E_J R)$ the relevant time scale of the circuit as detailed in App.~\ref{app:dimensionless} and the dimensionless time $\tau = t/t_0$. Here $E_J$ is the typical energy scale of the superconducting diode, defined as the difference between the maximum and the minimum of the Josephson potential. It enters the current unit $I_0 = 2e E_J/\hbar$, defining the dimensionless currents $\mathcal{I}(\varphi_1) \equiv I(\varphi_1)/I_0$ and $\mathcal{I}_\text{ext} \equiv I_\text{ext}/I_0$.

These definitions allow us to introduce the dimensionless potential as
\begin{equation}\label{eq:potential}
   \mathcal{U}(\varphi_1) = \int_{0}^{\varphi_1}\!\mathcal{I}(x)\,dx\,,
\end{equation}
with $\mathcal{U}_\text{max} - \mathcal{U}_\text{min} = 1$. Inserting those quantities into Eqs.~\eqref{eq:system}, the stochastic differential equations (SDEs) take the form
\begin{equation}
\label{eq:dimensionless}
\begin{aligned}
    \frac{d \varphi_1}{d\tau} &= -\mathcal{I}(\varphi_1) - \mathcal{I}_\text{ext} - \varkappa (\varphi_1 - \varphi_2) +  \sqrt{\frac{2T_1}{T_0}}\,\xi_1(\tau),\\
    \frac{d\varphi_2}{d\tau} &= \varkappa (\varphi_1 - \varphi_2) +  \sqrt{\frac{2T_2}{T_0}}\, \xi_2(\tau),
\end{aligned}
\end{equation}
with $T_0 = E_J/k_B$ and $\xi_j(\tau)$ normalized Gaussian white noises, with $\langle \xi_j (\tau)\rangle = 0$ and $\langle \xi_j(\tau) \xi_{j'}(\tau')\rangle = \delta(\tau -\tau')\delta_{j j'}$, with $j=1,2$.

When the two reservoirs have the same temperature,
$T_1=T_2\equiv T$, the system is effectively in contact with a single thermal bath. Since the Langevin equations satisfy the
fluctuation–dissipation relation, the dynamics obeys
detailed balance and relaxes to thermal equilibrium. The stationary probability density for the variables $\varphi_1, \varphi_2$ is then given by the Boltzmann distribution associated with the effective potential of the system $\mathcal{U}(\varphi_1)$,
\begin{equation}\label{eq:Prob.solution}
P(\varphi_1,\varphi_2) = 
\frac{\sqrt{\varkappa T_0}\,
\exp\!\left[-\frac{\mathcal{U}(\varphi_1)T_0}{T}
-\frac{\varkappa T_0}{2T}(\varphi_1-\varphi_2)^2\right]}
{\sqrt{2\pi T}\int_0^{2\pi}d\varphi_1
\exp\!\left[-\frac{\mathcal{U}(\varphi_1)T_0}{T}\right]},
\end{equation}
where the potential $\mathcal{U}(\varphi_1)$ is defined in Eq.~\eqref{eq:potential}, see App.~\ref{app:Probderivation}. Consequently, the probability current
vanishes and the average phase velocity $\langle \dot\varphi_1 \rangle=0$, implying zero voltage according to Eq.~\eqref{eq:volphase}.

\subsection{A Feynman ratchet under nonequilibrium conditions}
In the general case of two different temperatures, solving the Langevin equations requires knowledge of the exact CPR $\mathcal{I}(\varphi_1)$, which depends on the model chosen for the superconducting diode. In this work, we follow a model first introduced by Bozkurt et al.\ in Ref.~\cite{bozkurt2023double}, where the superconducting diode is designed as a network of Josephson tunnel junctions (JTJs). More specifically, we consider eight JTJs arranged into four parallel branches labeled $i=1,2,3,4$, see Fig.~\ref{fig:circuit}(b). In this situation, the CPR takes the form
\begin{equation}
\begin{gathered}
    I(\varphi_1) = \sum_{i=1}^4 \frac{\pi E_{J}^{(i)} \lambda_i}{2 \Phi_0} \frac{\sin(\varphi_1 - \phi_i)}{\sqrt{1 - \lambda_i \sin^2(\frac{\varphi_1 -\phi_i}{2})}},
\end{gathered}
\end{equation}
with $\Phi_0 = h/(2 e)$ the superconducting flux quantum. The parameters $E_J^{(i)}, \lambda_i$ and $\phi_i$ are defined in Table~\ref{tbl:parameters}.

To obtain the most effective Feynman ratchet, we require a diode with the largest possible rectification coefficient $\eta_d$, defined as
\begin{equation}
    \eta_d = \frac{|I_{c^+} - I_{c^-}|}{I_{c^+} + I_{c^-}} = \frac{|\mathcal{I}_{c^+} - \mathcal{I}_{c^-}|}{\mathcal{I}_{c^+} + \mathcal{I}_{c^-}},
\end{equation}
where $I_{c^\pm}$ are the critical currents for current flowing in the $\pm$ directions. The parameters are chosen to maximize $\eta_d$ in this configuration, corresponding to the values provided in Table~\ref{tbl:parameters}, leading to $\eta_d \approx 0.6$. We note that the rectification coefficient could be further increased at the cost of adding more branches to the model~\cite{bozkurt2023double}. Since this work does not aim to demonstrate the optimal operating regime of such a device, but rather proposes for the first time a Feynman ratchet based on a superconducting diode that can be exploited as a heat engine, we restrict ourselves to tractable numerical simulations corresponding to the case of four branches.

In the next section, we numerically investigate and characterize the voltage $V_1$ arising from the asymmetric CPR and the nonequilibrium situation, in the absence of an external current, $\mathcal{I}_\text{ext} = 0$.
\begin{table}[tb]
\centering
\begin{tabular}{cp{5mm}cp{5mm}cp{5mm}c}
    \hline\hline
    $i$& & $E_{J}^{(i)} /E_J$& & $\lambda_i $& & $ \phi_i $\\
     \hline
     1& &4.07& & 0.63 && 0.00\\
     2& &2.77& & 0.86 && 4.81\\
     3& &3.34& & 0.73& & 1.47\\
     4& &2.81& & 0.98& & 2.71\\
     \hline\hline
\end{tabular}
\caption{The parameters used in the circuit of Fig.~\ref{fig:circuit}(b).  They implement an (effective) Josephson diode with rectification coefficient $\eta_d \approx 0.6$.}
\label{tbl:parameters}
\end{table}

\section{Thermally induced voltage}\label{sec:VOLTAGE}

In this section, we investigate the thermo-voltage produced by the setup in Fig.~\ref{fig:circuit}(a) in the absence of an applied external current ($I_\text{ext}=\mathcal{I}_\text{ext} = 0$), under a thermal nonequilibrium situation with $T_1 \neq T_2$. We furthermore maximize the Seebeck coefficient to identify the optimal parameters for the strongest thermoelectric response.

\subsection{Open-circuit voltage}\label{sec:voltage}
As discussed in the previous section, the circuit we consider is characterized by three parameters: the coupling constant $\varkappa$ and the two dimensionless temperatures $T_1/T_0$ and $T_2/T_0$. The thermo-voltage $V_1$ generated by the circuit depends on these parameters, and since we are in the tight-coupling regime, the voltages in the two branches can be assumed to be approximately the same, $V_1 \approx V_2$, as discussed earlier.

\begin{figure}[tb]
    \centering
    \includegraphics{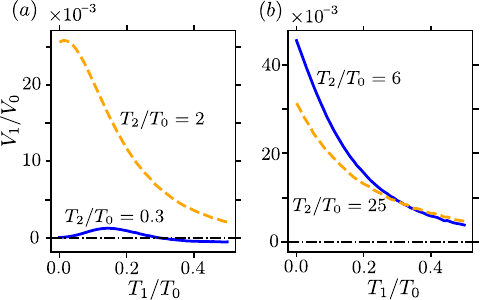}
\caption{Open-circuit voltage is shown in units of $V_0 = I_0R$, while the temperature $T_1$ of the `cold' reservoir is given in units of $T_0 = E_J/k_B$. The values of the `hot' reservoir temperature $T_2$ are indicated in the inset. The coupling strength is fixed at $\varkappa = \hbar^2 / (4 e^2 E_J L_{\text{eff}}) = 0.1$ for all curves. Panel (a) demonstrates that for $T_2 \lesssim 2.3T_0$, the maximum voltage occurs at finite $T_1$, whereas for $T_2 \gtrsim 2.3T_0$, the maximum is reached at $T_1 \to 0$ [panel (b)]. The blue curve for $T_2 = 6T_0$ and $T_1 = 0$ marks the overall maximum voltage $V_1 = 0.045V_0$ attainable at $\varkappa = 0.1$.}
    \label{fig:voltagetemp}
\end{figure}

With this assumption, we numerically solve the Langevin equations, Eqs.~\eqref{eq:dimensionless}, using Euler's method. In the presence of a thermal gradient between the two branches, the circuit develops a finite thermo-voltage, which changes sign depending on which reservoir is set to the hot or cold temperature. In the regime where $T_1<T_2$, the voltage is positive. We show this nonreciprocal behavior in Fig.~\ref{fig:voltagetemp}(a).

We first examine the maximum voltage achievable for $\varkappa = 0.1$. Figure~\ref{fig:voltagetemp} shows the voltage as a function of the cold-reservoir temperature $T_1$. For fixed $T_2 \lesssim 2.3T_0$, the maximum occurs within $0 < T_1 < T_2$, as shown by the blue curve in Fig.~\ref{fig:voltagetemp}(a). With increasing hot-reservoir temperature $T_2$, this maximum shifts toward lower $T_1$, reaching $T_1 = 0$ for $T_2 \gtrsim 2.3T_0$. For $\varkappa = 0.1$, the global open-circuit maximum, $V_1 = 0.045 V_0$, occurs at $T_2 = 6T_0$ and $T_1 = 0$, see Fig.~\ref{fig:voltagetemp}(b). For even larger temperatures $T_2$, the voltage decreases, while the maximum remains pinned at $T_1 = 0$, as illustrated by the dashed orange curve in Fig.~\ref{fig:voltagetemp}(b).

\begin{figure}[tb]
    \centering
    \includegraphics{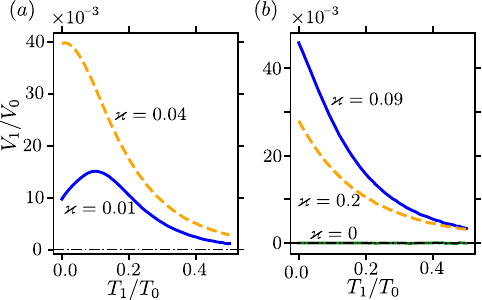}
    \caption{Open-circuit voltage is shown in units of $V_0 = RI_0$, and the cold-reservoir temperature $T_1$ is given in units of $T_0 = E_J/k_B$. The values of the dimensionless coupling constant $\varkappa$ are indicated in the inset. The hot-reservoir temperature is fixed at $T_2 = 5T_0$ for all curves. Panel (a) shows that for $\varkappa \lesssim 0.04$, the maximum voltage occurs at finite $T_1$, whereas for $\varkappa \gtrsim 0.04$ the maximum shifts to $T_1 \to 0$, see panel (b). For consistency, the case $\varkappa = 0$ is also included, where the voltage vanishes as expected.}
    \label{fig:voltagecoup} 
\end{figure}

Next, we fix the hot-reservoir temperature at $T_2 = 5T_0$ and vary both the coupling $\varkappa$ and the cold-reservoir temperature $T_1$. For weak coupling, $\varkappa \lesssim 0.04$, the maximal voltage obtained within $0 < T_1 < T_2$ is small, see Fig.~\ref{fig:voltagecoup}(a). As $\varkappa$ increases, the voltage grows and its maximum shifts toward lower $T_1$, reaching $T_1 \approx 0$ at $\varkappa = 0.04$, as indicated by the dashed orange curve in Fig.~\ref{fig:voltagecoup}(a). For stronger coupling, the maximum remains pinned at $T_1 = 0$, while the voltage continues to rise, reaching a global maximum of $V_1 = 0.046 V_0$ at $\varkappa = 0.09$. More generally, for small $\varkappa$, we find that the maximal open-circuit voltage occurs at $T_1 = 0$ and $T_2 \approx 0.5 T_0/\varkappa$.

We now investigate the thermoelectric properties of this superconducting diode through the Seebeck coefficient.

\subsection{Seebeck coefficient}\label{sec:Seebeck}
We emphasize that the external current $I_\text{ext}$ corresponds to a controllable bias current provided by an external source, as illustrated in Fig.~\ref{fig:volt_vs_curr}(b). It therefore acts as an externally imposed load on the device, with the resulting voltage determined by the thermoelectric response of the circuit. In the absence of an external load, $I_\text{ext}=0$, the device develops an open-circuit voltage across the junction as discussed in section~\ref{sec:voltage}. This voltage provides the natural starting point for characterizing the thermoelectric response in the linear-response regime, where the temperature difference $\Delta T \equiv T_2 - T_1$ is small. Therefore, the Seebeck coefficient is defined by
\begin{equation}\label{eq:Seebeck}
    S \equiv \frac{d V_1}{d \Delta T} \Bigg|_{\mathcal{I}_\text{ext}=0, \Delta T \to0}\,.
\end{equation}
The Seebeck coefficient can be expressed in terms of
\begin{equation}
    S_0 = \frac{RI_0}{T_0} = \frac{2 e}{\hbar} k_B R  \approx 42\,
    \mu\text{V/K}\,, 
\end{equation}
when considering a typical circuit resistance of $1\,\text{k}\Omega$. Note that the magnitude $S_0$ of the Seebeck coefficient only depends on $R$ and does not depend on the energy scale $E_J$ of the potential.

We compute $S$ numerically by plotting the voltage as a function of $T_1/T_0$ in a small window near equilibrium ($T_1 = T_2 = T$). By varying $\varkappa$ and $T$, we find a broad maximum at $\varkappa \approx 0.11$ and $T/T_0 \approx 0.24$. The maximal Seebeck coefficient is $S_\text{max} \approx 9.1\times 10^{-3} S_0$, corresponding to $S_\text{max} \approx 0.38\,\mu$V/K for $R = 1\,\text{k}\Omega$, and it increases linearly with resistance.

\section{Superconducting Feynman ratchet as heat engine and fridge}\label{sec:heatenginge}

In this section, we demonstrate how the circuit in Fig.~\ref{fig:circuit} can be operated as a heat engine exploiting its thermoelectric properties, and we analyze its performance. We conclude by discussing its reversed operating regime, acting as a fridge.

\subsection{Heat engine performance}\label{sec:efficiency}
\begin{figure}[tb]
    \centering
    \includegraphics{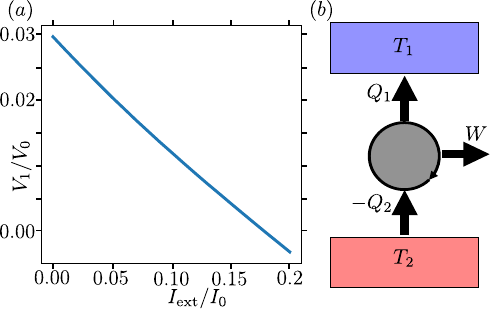}
    \caption{(a) Circuit voltage in units of $V_0 = R I_0$ as a function of $I_\text{ext}$ in units of $I_0$. The circuit parameters are $\varkappa = 0.1$, $T_2 = 2.3 T_0$ and $T_1 = 0$. This panel demonstrates that the dependence of the voltage as a function of the external applied current is linear in our regime. The point of maximum power is given for the values of $\mathcal{I}_\text{ext}$ in which the voltage takes half of its maximum value. (b) Heat flow when operated as a heat engine: the lower and upper branches of the circuit are presented respectively as hot and cold reservoirs. The circuit works as a heat engine, extracting the  heat $-Q_2$ from the hot reservoir, delivering  $Q_1$ to the cold reservoir while producing the work $W = -Q_1 -Q_2$. }
    \label{fig:volt_vs_curr}
\end{figure}

So far, we have considered the open-circuit voltage. For the system to function as a heat engine, however, it must deliver work, which requires attaching a load. We model this by introducing an external current $\mathcal{I}_\text{ext} > 0$ that extracts work in combination with the thermal voltage generated by the Feynman ratchet, see Eqs.~\eqref{eq:dimensionless}. Note that drawing a current reduces the voltage from its open-circuit value, and consequently there exists an optimal $\mathcal{I}_\text{ext}$ that maximizes the output power, $\dot{W} = V_1 I_0 \mathcal{I}_\text{ext}$. The value of $V_1$ in the presence of the external current is shown in Fig.~\ref{fig:volt_vs_curr}(a). For $\mathcal{I}_\text{ext} \geq 0$, it takes its maximum value at $\mathcal{I}_\text{ext}=0$ and then decreases as $\mathcal{I}_\text{ext}$ increases, vanishing at $\mathcal{I}_\text{ext} \approx 0.17$. Above that value, the voltage becomes negative. The maximum output power is reached at the value of $\mathcal{I}_\text{ext}$ for which the voltage is half its open-circuit value, where the product of voltage and current is largest.

\begin{figure}[tb]
    \centering
    \includegraphics{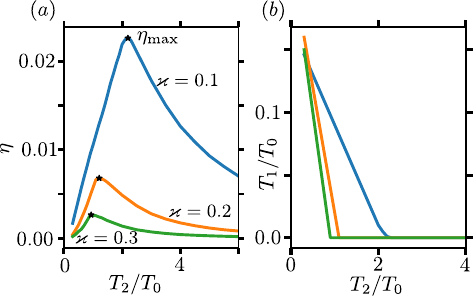}
    \caption{Panel (a) shows the heat-engine efficiency $\eta$ as a function of the hot-reservoir temperature $T_2$, given in units of $T_0$, for different values of the coupling strengths $\varkappa$. The maximal efficiency is denoted by $\eta_{\text{max}}$. As the coupling strength increases, the efficiency decreases. At the same time, the temperature $T_2$ at which the efficiency reaches its maximum shifts toward lower values. This behavior is further illustrated in panel (b), which shows the corresponding cold-reservoir temperature $T_1$ at maximum efficiency, also in units of $T_0$. For sufficiently large $T_1$, the dependence is approximately linear. As $T_1$ approaches zero, however, the curves become increasingly smooth, as exemplified by the blue curve. The maximum efficiency is attained in the limit $T_1 \gtrsim 0$. The maximal efficiency can be further increased by reducing the coupling strength, as shown in Fig.~\ref{fig:maxefficiency}.
    \label{fig:efficiency}}
\end{figure}

We emphasize that, for $\mathcal{I}_\text{ext} > 0$, the system operates as a heat engine, see Fig.~\ref{fig:volt_vs_curr}(b), absorbing heat $-Q_2$ from the hot reservoir, releasing heat $Q_1$ into the cold reservoir, and producing electrical work $W = -Q_1 - Q_2$. Following Ref.~\cite{sekimoto1998langevin}, we compute the work as\footnote{The stochastic integrals are evaluated using Stratonovich integration with a midpoint discretization.}
\begin{equation}\label{eq:work}
   \delta W = \frac{I_0 \hbar}{2 e} \mathcal{I}_\text{ext} d\varphi_1,
\end{equation}
and the heat as
\begin{equation}\label{eq:heat}
\delta Q_j = \frac{\hbar}{2 e}\left(\frac{\hbar}{2 e R} \dot{\varphi}_j - \delta I_j\right) d\varphi_j\,.
\end{equation}
It corresponds to the heat dissipated in the reservoir at temperature $T_j$ for a time-step that increases the phase by $d\varphi_j$.
In our simulations, we fix the external current $\mathcal{I}_\text{ext}$ and evolve the SDEs until the system reaches a steady state. We then compute the work output $\dot W$ and the heat currents $\dot Q_j$ per unit time. By the first law of thermodynamics, $\dot W + \dot Q_1 + \dot Q_2 = 0$, so knowledge of any two of these quantities determines the third. The efficiency is defined as $\eta = |\dot W / \dot Q_2|$ and is evaluated at maximum power as a function of the hot-reservoir temperature $T_2$, see Fig.~\ref{fig:efficiency}(a). For each $T_2$, the cold-reservoir temperature $T_1$ is chosen such that the open-circuit voltage is maximized, as depicted in Fig.~\ref{fig:efficiency}(b). We have plotted the efficiency, and the cold-reservoir temperature corresponding to maximum voltage for three different values of the coupling strength $\varkappa$.

The three curves in Fig.~\ref{fig:efficiency}(b) show that for low $T_2$, the cold-reservoir temperature $T_1$ corresponding to maximum voltage takes a finite value. As $T_2$ increases, the optimal $T_1$ decreases, while the efficiency $\eta$ increases. The maximum efficiency, $\eta_\text{max}$, occurs at vanishing $T_1$. For even larger hot reservoir temperatures, the efficiency decreases.

\begin{figure}[tb]
    \centering
    \includegraphics{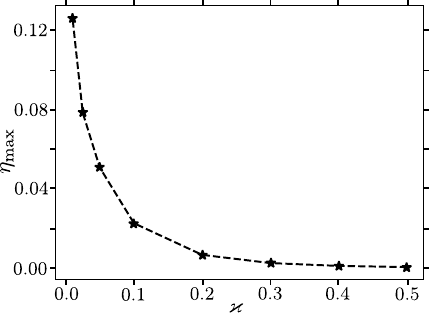}
    \caption{Maximum efficiency $\eta_\text{max}$ for different values of the coupling constant $\varkappa$. Each point presented corresponds to a different  temperature of the hot reservoir, with $\varkappa T_2 \approx 0.2 T_0$ and $T_1 = 0$. The line is a guide to the eye.}
    \label{fig:maxefficiency}
\end{figure}

Additional values of the maximum efficiency at different coupling strength are shown in Fig.~\ref{fig:maxefficiency}. We find that the efficiency increases as the coupling constant decreases. In particular, the maximum efficiency is obtained at $T_1 = 0$ and $T_2 \approx 0.2T_0/\varkappa$. For the smallest coupling constant considered, $\varkappa = 0.01$, we obtain $\eta_\text{max} \approx 12\%$ at $T_2 \approx 17T_0$. The maximum efficiency corresponding to $\varkappa = 0.1$ is $\eta_\text{max} \approx 2\%$.

\subsection{Peltier Cooling}\label{sec:Cooling}

While ratchet systems have mostly been investigated as heat engines~\cite{bang2018experimental}, they can also function as coolers~\cite{Broeck06, pekola07}.
In this section, we investigate whether our nano-heat engine can operate as a Peltier cooler. By reversing the external current  ($\mathcal{I}_\text{ext} < 0$), the signs of both heat currents $\dot Q_j$ are reversed, see Fig.~\ref{fig:volt_vs_curr}(b), allowing the system to act as a cooling device. Namely, by putting work into the circuit, the heat will flow from the cold reservoir to the hot reservoir. We expect the point of maximum Seebeck coefficient  to coincide with the optimal cooling regime in parameter space: at this point, a small temperature difference generates the largest voltage, which is necessary to drive the heat currents against their natural direction. This correspondence is consistent with the Thomson relation, which relates the Peltier and Seebeck coefficients.

\begin{figure}[tb]
    \centering
    \includegraphics{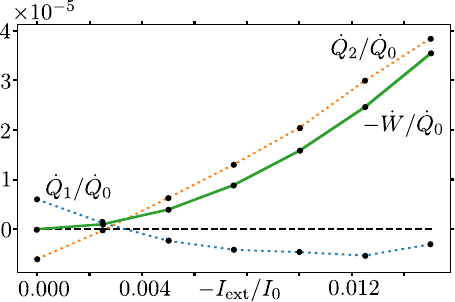}
    \caption{Heat and work (per unit of time) in the circuit when the upper branch is biased by an external current $I_{\text{ext}} < 0$ (reverse direction) in units of $\dot Q_0= 
    R I_0^2$. For $\mathcal{I}_{\text{ext}} = 0$, $\dot Q_2<0$, indicating that heat flows from the `hot' reservoir to the cold reservoir (and $\dot Q_1 = - \dot Q_2 >0$ according to the first law, see Fig.~\ref{fig:volt_vs_curr}(b)). Applying a sufficiently strong external current inverts the signs of the heat flows, as seen for $-\mathcal{I}_\text{ext}\gtrsim 0.003$. In this regime, the external current performs work on the circuit, cooling the cold reservoir.}
    \label{fig:Cooling}
\end{figure}

We study the system at the optimal parameters for the Seebeck coefficient, $\varkappa = 0.11$ and $T_1 \approx T_2 = 0.24 T_0$, with a small temperature difference applied. The required external current is determined by plotting the heat currents of both reservoirs and the work as a function of $\mathcal{I}_\text{ext}$. We find that for $-\mathcal{I}_\text{ext} \gtrsim 0.003$, the heat flow is reversed, confirming the operation of the system as a Peltier cooler, see Fig.~\ref{fig:Cooling}.

\section{Conclusions}\label{sec:conclusion}
In this paper,  we proposed a superconducting circuit realizing Feynman's ratchet. The circuit consists of two branches thermalized at different temperatures, with a superconducting diode providing non-reciprocity and acting as the ratchet. We demonstrated that the system can function both as a heat engine and as a Peltier cooler, achieving efficiencies on the order of $10\,\%$. The circuit generates an open-circuit voltage under a thermal gradient, with a Seebeck coefficient of approximately $S\approx 0.4\,\mu$V/K. Although the phenomena studied here are essentially classical, the device can operate at arbitrarily low temperatures. While we have proposed a superconducting diode as the ratchet element shown in Fig.~\ref{fig:circuit}(b), the underlying mechanism is more general and does not rely on this specific realization. Any asymmetric current-phase relation can give rise to the same effect. The system implements a nano-heat engine in which a few degrees of freedom are coupled to two heat reservoirs, closely mirroring Feynman's original concept.

\section*{Data Availability}
The data that support the findings of this study were generated by numerical simulations. The source code and parameters used to generate the simulations is publicly available~\cite{zenodo}.

\section*{Acknowledgments}
We acknowledge funding by the Deutsche Forschungsgemeinschaft (DFG, German Research Foundation) under Germany’s Excellence Strategy Cluster of Excellence Matter and Light for Quantum Computing (ML4Q) EXC 2004/1 390534769.
GH acknowledges funding from NCCR SwissMAP and from the Fondation Sandoz through the FPFS-Monique-de-Meuron program for academic promotion. 

\appendix
\section{Derivation of the stochastic differential equations}\label{app:SDEderivation}
In this appendix, we derive Eqs.~\eqref{eq:system}. First, we describe the inductive coupling between the upper and lower branches. In general, the voltage $V_1$ (of the upper branch) and current $I_1$ (through the inductive branch)  are related to the respective values $V_2$, $I_2$ in the lower branch via the linear relation
\begin{align}
    \begin{bmatrix}  V_1 \\ V_2  \end{bmatrix} = \begin{bmatrix} L & M\\ M & L\end{bmatrix} \begin{bmatrix} \dot{I}_1 \\ \dot{I}_2\end{bmatrix},
\end{align}
where $L$ is the self-inductance and $M$ is the mutual inductance. Using Eq.~\eqref{eq:volphase} and inverting the relation, we obtain
\begin{equation}
    \begin{bmatrix} I_1 \\ I_2 \end{bmatrix} = \frac{\hbar}{2 e} \frac{1}{L^2 - M^2} \begin{bmatrix} L & -M \\ - M & L\end{bmatrix} \begin{bmatrix} \varphi_1 \\ \varphi_2\end{bmatrix}.
\end{equation}
We are interested in the tight-coupling limit with $ L_\text{eff} \ll M $ where we have $I_2 \approx \hbar(\varphi_2 - \varphi_1) /(2 e L_\text{eff})$ and $L_\text{eff} = 2(L - M)$.
Adding all the different currents in the lower circuit, using Kirchhoff's law, we obtain the equation
\begin{align}\label{eq:branch_l}
    \frac{\hbar}{2 e R} \dot{\varphi}_2 = \frac{\hbar}{2 e} \frac{1}{L_\text{eff}} (\varphi_1 - \varphi_2) + \delta I_2.
\end{align}
Here we added $\delta I_2$, which is a random contribution due to Johnson-Nyquist noise. Similarly, the upper circuit leads to the equation
\begin{align}\label{eq:branch_u}
    \frac{\hbar}{2 e R} \dot{\varphi}_1 + I(\varphi_1) + I_\text{ext} + \frac{\hbar}{2 e} \frac{1}{L_\text{eff}} (\varphi_1 - \varphi_2) = \delta I_1.
\end{align}
Equations \eqref{eq:branch_u} and \eqref{eq:branch_l} form the starting point of our analysis of the system in the overdamped regime.

\section{Dimensionless form of the stochastic differential equations}\label{app:dimensionless}
In this appendix, we bring the Eqs.~\eqref{eq:system} into a dimensionless form. In this appendix we assume that there is no external current. We choose the dimensions of time and current to be 
\begin{align}\label{Units of time}
    t_0 &= \frac{\hbar^2}{(2 e)^2} \frac{1}{E_J R} &\quad\text{and}\quad
    I_0 &= \frac{2 e}{\hbar} E_J.
\end{align}
Introducing the dimensionless variables $\tau= t/t_0$ and  $\mathcal{I} = I/I_0$ in Eqs.~\eqref{eq:system}, we obtain
\begin{equation}
\begin{split}
\frac{d \varphi_1}{d\tau} &= -  \frac{2 e}{\hbar} I_0 R t_0 \mathcal{I}(\varphi_1) - \frac{R t_0}{L_{\text{eff}}} (\varphi_1 - \varphi_2) + \frac{2 e}{\hbar} I_0 R t_0 \delta \mathcal{I}_1,\\
\frac{d \varphi_2}{d\tau} &= \frac{R t_0}{L_{\text{eff}}} (\varphi_1 - \varphi_2) + \frac{2 e}{\hbar} I_0 R t_0 \delta \mathcal{I}_2.
\end{split}
\end{equation}
In the above equation, all terms are dimensionless. We introduce the dimensionless coupling 
\begin{equation}
    \varkappa = \frac{R t_0}{L_{\text{eff}}} =\frac{\hbar^2}{(2 e)^2} \frac{1}{E_J L_{\text{eff}}}.
\end{equation}
This describes the inductance effects between the upper and lower branches of the circuit. With this, Eqs.~\eqref{eq:system} take the form of Eqs.~\eqref{eq:dimensionless}.

In the analysis, we also need the unit of voltage 
\begin{equation}
    V_0 = I_0 R = \frac{2 e}{\hbar} E_J R\,,
\end{equation}
the temperature
\begin{equation}
    T_0 = \frac{E_J}{k_B}\,,
\end{equation}
and the  Seebeck coefficient
\begin{equation}\label{untis of Seebeck coef}
    S_0 = V_0/T_0 = \frac{2 e}{\hbar} k_B R\,.
\end{equation}
Finally, the units for power are
\begin{equation}
    \dot{Q}_0 = \left(\frac{2 e}{\hbar}\right)^2 E_J^2 R.
\end{equation}

\section{Derivation of the probability distribution at equal temperatures}\label{app:Probderivation}
Here, we derive Eq.~\eqref{eq:Prob.solution}. We start from the Fokker-Planck equation of the system which has the form
\begin{align}
    \frac{\partial P(\bm{\varphi},\tau)}{\partial \tau} = - \bm{\nabla_{\varphi}} \cdot\bm{J},
\end{align}
where $\bm{\varphi} = (\varphi_1, \varphi_2)$, and $\bm{J} = (J_1, J_2)$ being the vectors describing both phases and the dimensionless current densities respectively. The current components have the form
\begin{equation}
\begin{split}
    J_1 &= -\left[\mathcal{I}(\varphi_1) + \varkappa (\varphi_1 - \varphi_2)\right] P(\bm{\varphi},\tau) - \frac{T}{T_0} \frac{\partial P(\bm{\varphi},\tau)}{\partial \varphi_1}\,,\\
    J_2 &= - \varkappa (\varphi_2 - \varphi_1) P(\bm{\varphi},\tau) - \frac{T}{T_0} \frac{\partial P(\bm{\varphi},\tau)}{\partial \varphi_2}.
\end{split}
\end{equation}
When the temperatures are equal, for the second law of thermodynamics to be satisfied there should be no work done on the system, and thus  $\bm{J} = 0$. By solving the equation $J_2 = 0$, we get
\begin{align}\label{P}
    P(\varphi_1, \varphi_2) = f(\varphi_1) e^{- \frac{\varkappa T_0}{2 T} ( \varphi_2^2-2 \varphi_1 \varphi_2)},
\end{align}
where $f(\varphi_1)$ is an arbitrary function of $\varphi_1$. By substituting this solution into the equation $J_1 = 0$, we find a simple ODE for $f(\varphi_1)$, and by solving it and substituting back into Eq.~\eqref{P}, we get
\begin{align*}
    P(\varphi_1, \varphi_2) = P_0 \exp \biggl[{-\frac{T_0}{T} \int \mathcal{I}(\varphi_1) \,d\varphi_1} \!-\!\frac{\varkappa T_0}{2 T} (\varphi_1 \!-\! \varphi_2)^2\biggr],
\end{align*}
where $P_0$ is a constant of integration. We can find $P_0$ from the normalization condition $\int_0^{2 \pi} d\varphi_1 \,\int_{-\infty}^{\infty} d\varphi_2 \, P(\varphi_1, \varphi_2) = 1$.

\bibliography{references}

\end{document}